# LINEAR COMPLEX SUSCEPTIBILITY OF LONG RANGE INTERACTING DIPOLES WITH THERMAL AGITATION AND WEAK EXTERNAL AC FIELDS

Pierre-Michel Déjardin[1], Sergey V. Titov[2] and Yann Cornaton[1]

[1] *Laboratoire de Mathématiques et Physique, Université de Perpignan Via Domitia, 52, Avenue Paul Alduy, F-66860 Perpignan*

[2] *Kotel'nikov Institute of Radio Engineering of the Russian Academy of Sciences, Vvedenskii Square 1, Fryazino, Moscow Region 141120, Russia*

## ABSTRACT

An analytical formula for the linear complex susceptibility of dipolar assemblies subjected to thermal agitation, long range interactions and an externally applied uniform sinusoidal field of weak amplitude is derived using the forced rotational diffusion equation of Cugliandolo et al. [Phys. Rev. E **91**, 032139 (2015)] in the virial approximation. If the Kirkwood correlation factor of the dipolar assembly $g_K$ exceeds unity, a thermally activated process arising from the interaction-specific component arises while for $g_K < 1$, the susceptibility spectrum normalized by its static value is practically unaltered with respect to that of the ideal gas phase.

# I. INTRODUCTION

Ensembles of interacting dipoles are a useful idealization of the physics involved when modelling the linear electric susceptibility of dense polar fluids and the magnetic susceptibility of assemblies of single-domain ferromagnetic particles. The theory was initiated by Debye [1], who calculated first the static susceptibility of an assembly of rigid polar molecules, idealized as point dipoles, obtaining a result which is essentially a replica of Langevin's theory of paramagnetism and so is called the Langevin-Debye theory. He then extended the calculation to include the linear dynamic dielectric susceptibility (and therefore linear dielectric relaxation) of polar molecules subjected to a weak AC electric field, which unlike the static situation presents a non-equilibrium problem. In order to accomplish this, he treated the effects of the heat bath surrounding a dipole via the rotational diffusion model, thereby generalizing Einstein's theory of translational Brownian motion. However, the Debye theory holds only when long range forces and torques are neglected.

Now, a comprehensive theory of the *static* dielectric permittivity $\varepsilon(0)$ has been given by Kirkwood [2] and Fröhlich [3]. This theory properly accounts for long range torques and renders $\varepsilon(0)$ as a function of the density of the specimen, the dipole moment of a molecule in the ideal gas phase, the absolute temperature $T$ and a parameter $g_K$, the Kirkwood correlation factor. Loosely speaking, the value of $g_K$ accounts for dipolar order in the substance at equilibrium. If $g_K > 1$, the dipoles have a trend to orient parallel, if $g_K < 1$, they have a trend to orient antiparallel and if $g_K = 1$ no orientation is preferred *although the molecules interact* via long range forces and torques. The calculation of the frequency-dependent linear permittivity $\varepsilon(\omega)$ of dense (isotropic) polar liquids poses a much more involved theoretical problem because a) the dynamics of the field seen by a typical molecule in the Kirkwood-Fröhlich cavity is generally unknown, and its relation to the AC Maxwell field is not easy to establish for all AC field strengths [4], b) the account of dynamical correlations that would allow an *exact* microscopic calculation of the dynamical polarization has yet to be achieved.

The first author attempting a calculation of the dynamic susceptibility of interacting dipoles from basic principles was Zwanzig [5]. In the context of dielectrics, he considered electric dipoles (each representing the polarization state of a molecule) at the sites of a simple cubic lattice which in turn altogether constitutes a spherical sample in vacuum. He then used perturbation theory to calculate the polarizability of a sample to demonstrate that long-range forces were responsible for a microscopic, *discrete*, distribution of relaxation times. The overall relevant *macroscopic* relaxation time was found to *increase* when increasing the amplitude of intermolecular torques. However, Zwanzig's calculations are not amenable to comparison with experiment in practice, mostly because the perturbation theory calculations involved were too restrictive to be exploited for a dense system of molecules.

A different picture of dielectric relaxation was proposed by Nee and Zwanzig [6], based on dielectric friction, a memory mechanism that was introduced earlier by Zwanzig [7], its intention being to generalize Onsager's theory to the dynamical regime, objecting to the Onsager-Cole equation [8]. Proceeding, Nee and Zwanzig were able to reproduce the Fatuzzo-Mason equation for the frequency-dependent complex permittivity $\varepsilon(\omega)$ which is [9]

$$\frac{\varepsilon(0)\left[\varepsilon(\omega)-1\right]\left[2\varepsilon(\omega)+1\right]}{\varepsilon(\omega)\left[\varepsilon(0)-1\right]\left[2\varepsilon(0)+1\right]} = \left\{1+i\omega\tau + \frac{\left[\varepsilon(0)-1\right]\left[\varepsilon(0)-\varepsilon(\omega)\right]}{\varepsilon(0)\left[2\varepsilon(\omega)+1\right]}\right\}^{-1}$$

where $\tau$ is a relaxation time. The left hand side of this equation was later related by Fatuzzo and Mason in a later paper [9] to microscopic processes in an empirical manner. Nee and Zwanzig were further able to interpret quantitatively dielectric relaxation data of glycerol at – 60°C. However, their approach is subjected to criticism [10] in so far as it is not easy to understand how *short range torques* can combine with dielectric friction. Moreover, the theory cannot be connected to the static values predicted by the Kirkwood-Fröhlich equation because it completely neglects orientational correlations.

Yet another contribution to the subject was made by Madden and Kivelson [11], who attempted to generalize the Kirkwood-Fröhlich equation to dynamics. They demonstrated that 3 extraneous *dynamical* parameters in addition to static ones must be included in order to describe linear dielectric relaxation by curve fitting. They further proposed a relation between microscopic and macroscopic quantities, following a line of reasoning that was earlier proposed by Fulton [12]. Their theory was successfully compared with dielectric relaxation of $CH_2Cl_2$ at room temperature, taking a Kirkwood correlation factor of unity. However, no other temperature was considered, so that the temperature variation of their 3 dynamical parameters is unknown, and may be difficult to establish. Yet another formalism was proposed by Bagchi and Chandra, who calculated the space and time-dependent linear dielectric function using a generalized hydrodynamic formalism [13]. For practical purposes, these authors confined themselves to a particular type of dynamical random-phase approximation [14], where the interaction potential to thermal energy ratio is replaced by the direct roto-translational pair correlation function. Thus, their explicit results are confined to very dilute systems. Finally, the most complete analytical equation for the linear complex permittivity that can be obtained using linear response theory was derived by Scaife [4]. Yet, in order to use Scaife's formula in practice, a model for microscopic dynamics *has to be specified*.

For single-domain ferromagnetic particles, the theoretical subject was initiated by Néel [15]. In effect, Néel argued that the magnetic stability of fine magnetic particles was the manifestation of *apparent magnetic states*. More precisely, if the time of measurement of the (non-interacting) magnetic particles is shorter than a certain time scale (noted $\tau_{ref}$ in this paragraph for convenience), then the apparent magnetic state is a stable one (*ferromagnetic* or *blocked state*) while if the time of measurement is larger than $\tau_{ref}$, then the apparent magnetic state is paramagnetic (*superparamagnetic state*) as a result of thermal fluctuations that randomize the direction of the magnetic moment of a typical particle. Then he proposed a formula [15] for $\tau_{ref}$ which must have Arrhenius form at low temperatures, since any apparently stable magnetic state is associated with one of the minima of the magnetic free energy of a particle, and thermal agitation may cause spontaneous jumps from one state to the other. With this picture in mind, for an energy landscape of a particle having minima at the north and south poles of the unit sphere, $\tau_{ref}$ is called the time of reversal, or equally well the thermally activated time scale of magnetic nanoparticles, or the Néel time. The equation "*time of measurement = $\tau_{ref}$*" defines what is commonly called the *blocking temperature of the particle*. From what precedes, it is clear that $\tau_{ref}$ also describes the duration of magnetic stability of a particle. Yet Néel's theory for $\tau_{ref}$ lacked a derivation from first principles. The theory of magnetization reversal under thermal agitation was later put on a firm theoretical basis by Brown [16]. By emphasizing the need of a dynamical theory, he set up the Langevin equation for the magnetization of a tagged magnetic nanoparticle, treating the various energies involved as arising from external causes and the magnetization as a single mechanical entity. By deriving the Fokker-Planck equation (FPE) corresponding to his Langevin equation, he used both the Kramers treatment of thermally activated chemical reaction rates by diffusion processes [17] and a constrained variational method applied to the calculation of the smallest non-zero eigenvalue $\Lambda_1$ of the FPE [18] in order to evaluate $\tau_{ref} = \Lambda_1^{-1}$, demonstrating that in the overdamped case, both routes are equivalent and lead to an Arrhenius-Kramers-like thermally activated process for magnetization reversal. The theoretical range of validity of the so-called Néel-Brown formula has been discussed in many occasions in a more precise way than in Brown's qualitative discussion [19]. Nevertheless, like in the Debye theory, Brown's treatment ignores the effect of long range inter-particle interactions.

The importance of long range interactions in assemblies of magnetic nanoparticles is commonly admitted nowadays [20, 21], contiguously with that of particle size distributions which have also a trend to broaden the magnetic low-frequency absorption spectrum [22]. The relevant relaxation time was believed to increase with the interaction strength in a very large majority of systems [20]. This idea was first theoretically put forward by Shtrikman and Wohlfarth [23] who made simple mean field-like considerations to justify the Vogel-Fulcher-Tamman law for the relaxation time of nanoparticle assemblies. This idea was also developed by Dormann, Bessais and Fiorani [24] who proposed an empirical model for the relaxation time treating the interaction field as a mean field. This model also predicts an increase of the median blocking temperature (maximum in the peak of the zero field cooled magnetization – and therefore, the temperature peak of the magnetic susceptibility) of the assembly. Many numerical simulations concerning this problem have been achieved using various techniques, which at times agree with this "law" of relaxation time increase (see for example [25]), and at times are at variance (see for example [26]).

As a matter of fact, very few analytical approaches which start from first principles exist. We mention here the approach of Zubarev and Iskakova [27] who obtained very complicated expressions for the resulting susceptibility. Restricting themselves to a special orientational static pair distribution function, in effect leading to a kind of mean field approximation, Ivanov and Kuznetzova [28] obtained an analytical formula for the static susceptibility. These calculations were extended to the dynamical regime and have been checked against numerical simulations and experimental data [29] only recently. Yet, restricting themselves to a mean field-like approximation, orientational correlations have been neglected. One of us calculated the thermally activated time scale of magnetic relaxation of magnetic nanoparticle assemblies by suitably adapting Berne's theory [30] of dielectric relaxation to magnetic particles having finite uniaxial anisotropy [31], with a result for the reversal time that is similar to that of Zwanzig [5]. However, although demonstrating an increase of the thermally activated time scale while increasing the concentration of the particles, this last approach is valid for very low densities only, because it has a Curie point as a consequence of ignoring inter-particle correlations. In summary, the starting point of all these models are special cases of the Dean-Kawasaki formalism [32] described by Cugliandolo et al. for dipolar dynamics [33].

Returning to dielectrics, an analytical calculation of $g_K$ in terms of dipole moment, density and temperature has been proposed quite recently using the Dean-Kawasaki formalism adapted to rotational Brownian motion of interacting dipoles [34]. This elegant formalism allows generalization of the Debye theory to straightforwardly include intermolecular interactions in the collective dipolar dynamics and can be solved *analytically* in the virial approximation. Furthermore, it yields excellent agreement with experimental data on the static linear dielectric constant vs temperature for many simple polar liquids, including liquid water and liquid methanol, showing that in the static regime, the microstructure is mostly concealed in the density of the liquid under study, and that distortional polarization effects may not be completely separated in the Kirkwood-Fröhlich equation of state [34]. Nevertheless, the results obtained in [34] are restricted to pure statics.

The purpose of this work is to obtain an analytical formula for the polarizability $\alpha(\omega)$ of a system of long range interacting dipoles via the Dean-Kawasaki rotational equation of Cugliandolo et al. [33], in which orientational correlations are no longer neglected, in order to extend the results of Reference 34 valid for statics. The language of dielectrics will be used throughout this work, but the calculations presented below can readily be transposed to magnetic relaxation of fine particles with zero anisotropy, in particular blocked ferrofluids and mechanically frozen magnetic nanoparticle assemblies with zero individual anisotropy. The derivation will start by stating the averaged Dean-Kawasaki equations for the one-body and pair orientational density, showing the rise of a Bogolyubov-Born-Green-Kirkwood-Yvon (BBGKY) like hierarchy. Then, a procedure for stopping the BBGKY process at the pair density level will be proposed, and an analytical formula for the dynamic susceptibility of the assembly of interacting dipoles will be derived.

## II.  MICROSCOPIC DIPOLAR DYNAMICS

We consider throughout this work a sample of volume $\Upsilon$ made of polar molecules (of a single species) each of which carries a dipole moment vector $\boldsymbol{\mu}$. These molecules are coupled via long-range intermolecular interactions. We shall further assume that such molecules are fixed in space, but can rotate at their sites. The total dipole moment per unit volume $P(t)$ at time $t$ in the direction of an applied field is given by the equation

$$P(t) = \rho_0 \mu \int (\mathbf{u} \cdot \mathbf{e}) W(\mathbf{u},t) d^2\mathbf{u}, \qquad (1)$$

where $\mathbf{e}$ denotes the direction of the applied field, $\mathbf{u}$ is a unit vector along a tagged dipole moment, $\rho_0$ is the number of molecules per unit volume of the sample and $W(\mathbf{u},t)$ is the probability density of orientations of individual dipole moments on the unit sphere. The computation of $P(t)$ by microscopic means essentially requires the knowledge of the dynamics of $W(\mathbf{u},t)$. We choose this dynamics to be governed by the averaged rotational Dean-Kawasaki equation [33], viz.

$$2\tau_D \frac{\partial W}{\partial t}(\mathbf{u},t) = \nabla_\mathbf{u} \cdot \left[ \nabla_\mathbf{u} W(\mathbf{u},t) + \beta W(\mathbf{u},t) \nabla_\mathbf{u} V_1(\mathbf{u},t) \right]$$
$$+ \beta \nabla_\mathbf{u} \cdot \int \nabla_\mathbf{u} U_m(\mathbf{u},\mathbf{u}') W_2(\mathbf{u},\mathbf{u}',t) d\mathbf{u}' \qquad (2)$$

where $\beta = (kT)^{-1}$, $k$ is Boltzmann's constant, $T$ is the absolute temperature, $\tau_D = \beta \varsigma / 2$ is the rotational diffusion time, $\varsigma$ is a phenomenological rotational friction coefficient, $\nabla_{\mathbf{u}}$ is the del operator on the unit sphere, $V_1(\mathbf{u},t)$ is a single molecule potential that may be written

$$V_1(\mathbf{u},t) = -\mu \mathbf{F}_\ell(t) \cdot \mathbf{u}, \tag{3}$$

$\mathbf{F}_\ell(t)$ is the time-dependent internal field (assumed uniform), $U_m(\mathbf{u},\mathbf{u}')$ is the (long range) pair interaction potential and $W_2(\mathbf{u},\mathbf{u}',t)$ is the orientational pair probability density. It has to be noticed that in writing Eq. (2), we restricted ourselves to pair interactions only.

As it stands, Eq. (2) is a rotational Smoluchowski equation forced by pair interactions. In order to solve it, one requires an equation for $W_2(\mathbf{u},\mathbf{u}',t)$. Using the formalism provided in Reference 33, it may be shown that this equation is

$$\begin{aligned}2\tau_D \frac{\partial W_2}{\partial t}(\mathbf{u},\mathbf{u}',t) &= \nabla_{\mathbf{u}} \cdot \left[ \nabla_{\mathbf{u}} W_2(\mathbf{u},\mathbf{u}',t) + \beta W_2(\mathbf{u},\mathbf{u}',t) \nabla_{\mathbf{u}} V_2(\mathbf{u},\mathbf{u}',t) \right] \\ &+ \nabla_{\mathbf{u}'} \cdot \left[ \nabla_{\mathbf{u}'} W_2(\mathbf{u},\mathbf{u}',t) + \beta W_2(\mathbf{u},\mathbf{u}',t) \nabla_{\mathbf{u}'} V_2(\mathbf{u},\mathbf{u}',t) \right] \\ &+ \beta \nabla_{\mathbf{u}} \cdot \int \nabla_{\mathbf{u}} U_m(\mathbf{u},\mathbf{u}'') W_3(\mathbf{u},\mathbf{u}',\mathbf{u}'',t) d\mathbf{u}'' \\ &+ \beta \nabla_{\mathbf{u}'} \cdot \int \nabla_{\mathbf{u}'} U_m(\mathbf{u}',\mathbf{u}'') W_3(\mathbf{u},\mathbf{u}',\mathbf{u}'',t) d\mathbf{u}''\end{aligned}$$

(4)

where $W_3(\mathbf{u},\mathbf{u}',\mathbf{u}'',t)$ is the three-body orientational probability density and

$$V_2(\mathbf{u},\mathbf{u}',t) = -\mu \mathbf{F}_\ell(t) \cdot (\mathbf{u}+\mathbf{u}') + U_m(\mathbf{u},\mathbf{u}') \tag{5}$$

is a pair potential. The solution of Eq. (4) in turn requires an equation for $W_3$, which will involve the four-body probability density $W_4$ and so on. Thus, the Dean-Kawasaki formalism generates a hierarchy of differential-integral equations that are similar to the Bogolyubov-Born-Green-Kirkwood-Yvon (BBGKY) hierarchy for partial densities. As such, there is no known systematic way of truncating this hierarchy [35] because stopping the BBGKY process by setting $W_{n+1} \equiv 0$ at the $n^{th}$ rank *is an incorrect procedure*. Hence, one must resort to non-trivial approximation schemes to estimate the effect of $W_{n+1}$ at the $n^{th}$ rank. We describe below a way to estimate $W_3$ in Eq. (4) so as to roughly describe orientational dipolar relaxation beyond the mean field approximation.

### III. REDUCTION OF EQ. (4) TO A TWO-BODY FOKKER-PLANCK EQUATION

Here, we show how to transform Eq. (4) into a two-body rotational Smoluchowski equation with an effective two-body potential. To achieve this, we first remark that $W_2$ and $W_3$ can generally be written as

$$W_2(\mathbf{u},\mathbf{u}',t) = W(\mathbf{u},t)W(\mathbf{u}',t)g(\mathbf{u},\mathbf{u}',t), \tag{6}$$

$$W_3(\mathbf{u},\mathbf{u}',\mathbf{u}'',t) = W(\mathbf{u},t)W(\mathbf{u}',t)W(\mathbf{u}'',t)g_3(\mathbf{u},\mathbf{u}',\mathbf{u}'',t), \tag{7}$$

where $g$ and $g_3$ are the orientational pair and three-body distribution functions respectively, which are taken as time-dependent. This allows, using Eqs. (6) and (7), to write $W_3$ in terms of $W_2$ as follows

$$W_3(\mathbf{u},\mathbf{u}',\mathbf{u}'',t) = W_2(\mathbf{u},\mathbf{u}',t)\frac{g_3(\mathbf{u},\mathbf{u}',\mathbf{u}'',t)}{g(\mathbf{u},\mathbf{u}',t)}W(\mathbf{u}'',t). \tag{8}$$

This last equation allows us to write Eq. (4) as

$$2\tau_D \frac{\partial W_2}{\partial t}(\mathbf{u},\mathbf{u}',t) = \nabla_\mathbf{u} \cdot \left[\nabla_\mathbf{u} W_2(\mathbf{u},\mathbf{u}',t) + \beta W_2(\mathbf{u},\mathbf{u}',t)\nabla_\mathbf{u} V_2^{eff}(\mathbf{u},\mathbf{u}',t)\right]$$
$$+ \nabla_{\mathbf{u}'} \cdot \left[\nabla_{\mathbf{u}'} W_2(\mathbf{u},\mathbf{u}',t) + \beta W_2(\mathbf{u},\mathbf{u}',t)\nabla_{\mathbf{u}'} V_2^{eff}(\mathbf{u},\mathbf{u}',t)\right] \tag{9}$$

where we have introduced a two-body time-dependent effective potential $V_2^{eff}(\mathbf{u},\mathbf{u}',t)$ via the partial differential equations

$$\nabla_\mathbf{u} V_2^{eff}(\mathbf{u},\mathbf{u}',t) = \nabla_\mathbf{u} V_2(\mathbf{u},\mathbf{u}',t) + \frac{1}{g(\mathbf{u},\mathbf{u}',t)}\int \nabla_\mathbf{u} U_m(\mathbf{u},\mathbf{u}'') W(\mathbf{u}'',t) g_3(\mathbf{u},\mathbf{u}',\mathbf{u}'',t) d\mathbf{u}'', \qquad (10)$$

$$\nabla_{\mathbf{u}'} V_2^{eff}(\mathbf{u},\mathbf{u}',t) = \nabla_{\mathbf{u}'} V_2(\mathbf{u},\mathbf{u}',t) + \frac{1}{g(\mathbf{u},\mathbf{u}',t)}\int \nabla_{\mathbf{u}'} U_m(\mathbf{u}',\mathbf{u}'') W(\mathbf{u}'',t) g_3(\mathbf{u},\mathbf{u}',\mathbf{u}'',t) d\mathbf{u}'', \qquad (11)$$

In its static version, $V_2^{eff}(\mathbf{u},\mathbf{u}',t)$ is analogous to the "potential of mean force" introduced earlier by Kirkwood [36] for translational degrees of freedom. In order to remove the first term in the right hand sides of Eqs. (10) and (11), we seek $V_2^{eff}(\mathbf{u},\mathbf{u}',t)$ in the form

$$V_2^{eff}(\mathbf{u},\mathbf{u}',t) = V_2(\mathbf{u},\mathbf{u}',t) + V_c(\mathbf{u},\mathbf{u}',t) \qquad (12)$$

where we term $V_c(\mathbf{u},\mathbf{u}',t)$ the "effective complementary potential" term, as it roughly contains the effects due to three-body correlations. Furthermore, Eq. (12) allows the elimination of external field terms contained in $V_2$. Using Eq. (12), Eqs. (10) and (11) become

$$\nabla_\mathbf{u} V_c(\mathbf{u},\mathbf{u}',t) = \frac{1}{g(\mathbf{u},\mathbf{u}',t)}\int \nabla_\mathbf{u} U_m(\mathbf{u},\mathbf{u}'') W(\mathbf{u}'',t) g_3(\mathbf{u},\mathbf{u}',\mathbf{u}'',t) d\mathbf{u}'', \qquad (13)$$

$$\nabla_{\mathbf{u}'} V_c(\mathbf{u},\mathbf{u}',t) = \frac{1}{g(\mathbf{u},\mathbf{u}',t)}\int \nabla_{\mathbf{u}'} U_m(\mathbf{u}',\mathbf{u}'') W(\mathbf{u}'',t) g_3(\mathbf{u},\mathbf{u}',\mathbf{u}'',t) d\mathbf{u}'', \qquad (14)$$

Now, since $V_c(\mathbf{u},\mathbf{u}',t)$ contributes to intermolecular interactions only, its time dependence is only parametric and can be neglected in Eqs. (13) and (14) *in a first approximation.* In other words, we ignore memory effects due to three-body correlations. This argument is reinforced by the fact that if we exclude external field terms, $V_2^{eff}(\mathbf{u},\mathbf{u}',t)$, should represent electrostatic or magnetostatic interactions, in which the time does not appear explicitly. Thus, Eqs. (13) and (14) become

$$\nabla_{\mathbf{u}} V_c(\mathbf{u},\mathbf{u}') = \frac{1}{g(\mathbf{u},\mathbf{u}')} \int \nabla_{\mathbf{u}} U_m(\mathbf{u},\mathbf{u}'') W(\mathbf{u}'') g_3(\mathbf{u},\mathbf{u}',\mathbf{u}'') d\mathbf{u}'' \qquad (15)$$

$$\nabla_{\mathbf{u}'} V_c(\mathbf{u},\mathbf{u}') = \frac{1}{g(\mathbf{u},\mathbf{u}')} \int \nabla_{\mathbf{u}'} U_m(\mathbf{u}',\mathbf{u}'') W(\mathbf{u}'') g_3(\mathbf{u},\mathbf{u}',\mathbf{u}'') d\mathbf{u}'' \qquad (16)$$

These two equations are similar in form to Eq. (15) of Rice and Lekner [37], the difference being that $W$ is non-uniform here and that the present work is devoted to rotational motion rather than translational and intermolecular vibrational ones.

It is evident that the statistical ensemble associated with $g_3(\mathbf{u},\mathbf{u}',\mathbf{u}'')$ is made of representative samples consisting of three molecules each carrying a single dipole. The three dipoles in a representative sample have orientations $(\mathbf{u},\mathbf{u}',\mathbf{u}'')$. We proceed by introducing the Kirkwood superposition approximation, which means that $g_3$ is simply the exponential of minus $\beta$ times the three-body interaction (this is the virial approximation for three bodies), and that the three-body interaction is made of superposition of pair interaction terms. Furthermore, since in the virial approximation, $g(\mathbf{u},\mathbf{u}') \approx e^{-\beta U_m(\mathbf{u},\mathbf{u}')}$, we necessarily have

$$g_3(\mathbf{u},\mathbf{u}',\mathbf{u}'') \approx g(\mathbf{u},\mathbf{u}') g(\mathbf{u}',\mathbf{u}'') g(\mathbf{u},\mathbf{u}'') \qquad (17)$$

We note in passing that Eq. (17) is exact for the state of maximal Boltzmann-Shannon entropy for a statistical ensemble made of representative samples composed of three interacting bodies [38]. On using Eq. (17) in conjunction with Eqs. (15) and (16), we have

$$\nabla_{\mathbf{u}} V_c(\mathbf{u},\mathbf{u}') \approx \int \nabla_{\mathbf{u}} U_m(\mathbf{u},\mathbf{u}'') W(\mathbf{u}'') g(\mathbf{u}',\mathbf{u}'') g(\mathbf{u},\mathbf{u}'') d\mathbf{u}'' \qquad (18)$$

$$\nabla_{\mathbf{u}}V_c(\mathbf{u},\mathbf{u}') \approx \int \nabla_{\mathbf{u}'}U_m(\mathbf{u}',\mathbf{u}'')W(\mathbf{u}'')g(\mathbf{u}',\mathbf{u}'')g(\mathbf{u},\mathbf{u}'')d\mathbf{u}'' \tag{19}$$

Now, we assume that a representative sample $(\mathbf{u},\mathbf{u}',\mathbf{u}'')$ is such that the dipole having orientation $\mathbf{u}''$ is *far away from the pair of dipoles with orientations* $(\mathbf{u},\mathbf{u}')$. To the lowest order of approximation, this results in

$$g(\mathbf{u}',\mathbf{u}'') \approx 1, \ g(\mathbf{u},\mathbf{u}'') \approx 1. \tag{20}$$

Thus, Eqs. (18) and (19) using Eqs. (20) become

$$\nabla_{\mathbf{u}}V_c(\mathbf{u},\mathbf{u}') \approx \int \nabla_{\mathbf{u}}U_m(\mathbf{u},\mathbf{u}'')W(\mathbf{u}'')d\mathbf{u}'', \tag{21}$$

$$\nabla_{\mathbf{u}'}V_c(\mathbf{u},\mathbf{u}') \approx \int \nabla_{\mathbf{u}'}U_m(\mathbf{u}',\mathbf{u}'')W(\mathbf{u}'')d\mathbf{u}'' \tag{22}$$

These two equations now allow us to seek $V_c(\mathbf{u},\mathbf{u}')$ as a superposition of single-body terms, namely

$$V_c(\mathbf{u},\mathbf{u}') = U_{an}(\mathbf{u}) + U_{an}(\mathbf{u}'), \tag{23}$$

where $U_{an}(\mathbf{u})$ is an even function of $\mathbf{u}$ on the unit sphere. Hence, Eqs. (21) and (22) become one and the same partial differential equation, viz.

$$\nabla_{\mathbf{u}}U_{an}(\mathbf{u}) \approx \int \nabla_{\mathbf{u}}U_m(\mathbf{u},\mathbf{u}'')W(\mathbf{u}'')d\mathbf{u}'', \tag{24}$$

We note here that $W(\mathbf{u}'')$ cannot be chosen as an equilibrium one-body probability density, because this would cause the right hand sides of Eqs. (24) to vanish ; this is similar to ignoring the integral terms in Eq. (4). This is however not permitted, because this would break the BBGKY-like nature of the averaged Dean-Kawasaki equations.

Now, we need to make a statement regarding $W(\mathbf{u}'')$ in Eq. (24). We set $\mathbf{u}=\mathbf{u}''$ in Eq. (2), write Eq. (6) as

$$W_2(\mathbf{u},\mathbf{u}',t) \approx W(\mathbf{u},t)W(\mathbf{u}',t)g(\mathbf{u},\mathbf{u}'), \qquad (25)$$

so that in the absence of external field, Eq. (2) with $\mathbf{u}=\mathbf{u}''$ is

$$2\tau_D \frac{\partial W}{\partial t}(\mathbf{u}'',t) = \nabla_\mathbf{u} \cdot \left[\nabla_\mathbf{u} W(\mathbf{u}'',t) + \beta W(\mathbf{u}'',t)\int \nabla_\mathbf{u} U_m(\mathbf{u}'',\mathbf{u}')W(\mathbf{u}',t)g(\mathbf{u}'',\mathbf{u}')d\mathbf{u}'\right] \qquad (26)$$

Hence, via Eq. (20), the integral term vanishes *to leading order* since as a consequence of the virial approximation for the orientational pair distribution function we have $U_m(\mathbf{u}'',\mathbf{u}') \approx 0$. Thus, instead of Eq. (26) we effectively have

$$2\tau_D \frac{\partial W}{\partial t}(\mathbf{u}'',t) \approx \nabla_\mathbf{u}^2 W(\mathbf{u}'',t), \qquad (27)$$

an equation that is meaningful only if we take (this is similar to Singer's point of view [38])

$$W(\mathbf{u}'',t) = \delta(\mathbf{u}''-\mathbf{u}) = W(\mathbf{u}''). \qquad (28)$$

Hence, Eq. (24) becomes effectively

$$\nabla_\mathbf{u} U_{an}(\mathbf{u}) \approx \nabla_\mathbf{u} U_m(\mathbf{u},\mathbf{u}'')\big|_{\mathbf{u}''=\mathbf{u}}, \qquad (29)$$

This last equation is the final result of the phase space reduction initiated at the start of this section. For the model pair interaction potential

$$\beta U_m(\mathbf{u},\mathbf{u}') = \mp\lambda\cos\vartheta\cos\vartheta', \qquad (30)$$

$\vartheta$ being the angle a dipole with orientation **u** makes with the local field $\mathbf{F}_\ell(t)$, $\lambda = 4\pi\beta\rho_0\mu^2/3$, the $\mp$ sign expressing the parallel ($g_K > 1$) or anti-parallel ($g_K < 1$) alignment of dipoles, Eq. (29) leads to ($z = \cos\vartheta$)

$$\beta \frac{dU_{an}}{dz}(z) = \mp \lambda z, \tag{31}$$

so that we have

$$\beta U_{an}(z) = \mp \frac{\lambda}{2} z^2. \tag{32}$$

From this equation and Eqs. (12), (23) and (30) we may rebuild the effective pair interaction potential $V_2^{eff}(z, z')$. We have four possible expressions given by all combinations of signs, viz.

$$\beta V_2^{eff}(z, z') = \begin{cases} -\dfrac{\lambda}{2}(z+z')^2 \\ -\dfrac{\lambda}{2}(z-z')^2 \\ \dfrac{\lambda}{2}(z+z')^2 \\ \dfrac{\lambda}{2}(z-z')^2 \end{cases} \tag{33}$$

Of these four expressions, the last two of them must be rejected because they lead to an artificial equilibrium position $(\vartheta, \vartheta') = (\pi/2, \pi/2)$ for a pair of dipoles, which is physically meaningless here. Hence, the two final effective pair interaction potentials corresponding to Eq. (30) are given by

$$\beta V_2^{eff}(z, z') = \begin{cases} -\dfrac{\lambda}{2}(z+z')^2 \\ -\dfrac{\lambda}{2}(z-z')^2 \end{cases}. \tag{34}$$

Finally, we may superpose externally applied fields onto Eq. (34) so that for the interaction potential given by Eq. (30), we have finally

$$\beta V_2^{eff}(\mathbf{u},\mathbf{u}',t) = -\frac{\lambda}{2}(\cos\vartheta \pm \cos\vartheta')^2 - \xi_\ell(t)(\cos\vartheta + \cos\vartheta'). \tag{35}$$

where $\xi_\ell(t)$ is a *time-dependent* reduced local field amplitude. This equation was used in its static version in Reference 34. Hence, Eqs. (9)-(11) reduce to the Fokker-Planck equation

$$2\tau_D \frac{\partial W_2}{\partial t}(\mathbf{u},\mathbf{u}',t) = \nabla_\mathbf{u} \cdot \left[\nabla_\mathbf{u} W_2(\mathbf{u},\mathbf{u}',t) + \beta W_2(\mathbf{u},\mathbf{u}',t)\nabla_\mathbf{u} V_2^{eff}(\mathbf{u},\mathbf{u}',t)\right] \\ + \nabla_{\mathbf{u}'} \cdot \left[\nabla_{\mathbf{u}'} W_2(\mathbf{u},\mathbf{u}',t) + \beta W_2(\mathbf{u},\mathbf{u}',t)\nabla_{\mathbf{u}'} V_2^{eff}(\mathbf{u},\mathbf{u}',t)\right] \tag{36}$$

with $V_2^{eff}(\mathbf{u},\mathbf{u}',t)$ now given by Eq. (35). We remark that Eq. (36) is similar to the dynamical equation that was given in Ref. 39, with the exceptions that $\xi_\ell(t)$ explicitly involves the dynamical internal field, that the Fokker-Planck equation is a two-body one and that the "anisotropy parameter" is expressed in terms of the molecular density and the individual dipole of a molecule.

Proceeding, we must compute the solution of Eq. (36) in the linear response regime, i.e. when $\xi_\ell(t) \ll 1$. This is the purpose of the next section.

## IV. MATRIX CONTINUED FRACTION SOLUTION OF THE FOKKER-PLANCK EQ. (36)

We may solve Eq. (36) starting from an expansion of $W_2(\mathbf{u},\mathbf{u}',t)$ in a series of Legendre polynomials $P_n(\cos\vartheta)$, viz.

$$W_2(\mathbf{u},\mathbf{u}',t) = \sum_{n=0}^{\infty}\sum_{m=0}^{\infty}\left(n+\frac{1}{2}\right)\left(m+\frac{1}{2}\right)b_{n,m}(t)P_n(\cos\vartheta)P_m(\cos\vartheta'), \tag{37}$$

and combine the recurrence and orthogonality of these polynomials with Eq. (36) to obtain the hierarchy of differential recurrence relations

$$\begin{aligned}
\tau_D \dot{b}_{nm}(t) &= q_{nm}^{(00)} b_{nm}(t) + q_{nm}^{(-20)} b_{n-2m}(t) + q_{nm}^{(20)} b_{n+2m}(t) \\
&+ q_{nm}^{(0-2)} b_{nm-2}(t) + q_{nm}^{(02)} b_{nm+2}(t) + q_{nm}^{(-1-1)} b_{n-1m-1}(t) + q_{nm}^{(11)} b_{n+1m+1}(t) \\
&+ q_{nm}^{(1-1)} b_{n+1,m-1}(t) + q_{nm}^{(-11)} b_{n-1,m+1}(t) \\
&+ \xi_\ell(t) \left[ p_{nm}^{(-10)} b_{n-1,m}(t) + p_{nm}^{(10)} b_{n+1,m}(t) + p_{nm}^{(0-1)} b_{n,m-1}(t) + p_{nm}^{(01)} b_{n,m+1}(t) \right]
\end{aligned} \qquad (38)$$

where the coefficients $q_{nm}^{(ij)}$ are specified in the Appendix. The set of Eqs. (38) may in turn be transformed into vector form, viz.

$$\tau_D \dot{\mathbf{c}}_n(t) = \mathbf{q}_n^{--} \mathbf{c}_{n-2}(t) + \mathbf{q}_n \mathbf{c}_n(t) + \mathbf{q}_n^{++} \mathbf{c}_{n+2}(t) + \xi_\ell(t) \left( \mathbf{p}_n^- \mathbf{c}_{n-1}(t) + \mathbf{p}_n^+ \mathbf{c}_{n+1}(t) \right), \qquad (39)$$

where

$$\mathbf{c}_n(t) = \begin{pmatrix} b_{n0}(t) \\ b_{n-11}(t) \\ \vdots \\ b_{0n}(t) \end{pmatrix}, \quad \mathbf{c}_0 = (1), \qquad (40)$$

and

$$\mathbf{q}_n = \begin{pmatrix}
q_{n0}^{(00)} & q_{n0}^{(-11)} & 0 & \ddots & 0 \\
q_{n-11}^{(1-1)} & q_{n-11}^{(00)} & q_{n-11}^{(-11)} & \ddots & \ddots \\
0 & q_{n-22}^{(1-1)} & q_{n-22}^{(00)} & \ddots & 0 \\
\ddots & \ddots & \ddots & \ddots & q_{1n-1}^{(-11)} \\
0 & \ddots & 0 & q_{0n}^{(1-1)} & q_{0n}^{(00)}
\end{pmatrix},$$

$$\mathbf{q}_n^{++} = \begin{pmatrix}
q_{n0}^{(20)} & q_{n0}^{(11)} & q_{n0}^{(02)} & 0 & \ddots & 0 & 0 \\
0 & q_{n-11}^{(20)} & q_{n-11}^{(11)} & q_{n-11}^{(02)} & 0 & \ddots & 0 \\
\ddots & 0 & q_{n-22}^{(20)} & q_{n-22}^{(11)} & q_{n-22}^{(02)} & \ddots & \ddots \\
0 & \ddots & \ddots & \ddots & \ddots & \ddots & 0 \\
0 & 0 & \ddots & 0 & q_{0n}^{(20)} & q_{0n}^{(11)} & q_{0n}^{(02)}
\end{pmatrix},$$

$$\mathbf{q}_n^{--} = \begin{pmatrix} q_{n0}^{(-20)} & \ddots & & & & 0 \\ q_{n-11}^{(-1-1)} & \ddots & \ddots & & & \\ q_{n-22}^{(0-2)} & \ddots & q_{2n-2}^{(-20)} & & \\ & \ddots & \ddots & q_{1n-1}^{(-1-1)} & \\ 0 & \ddots & q_{0n}^{(0-2)} & \end{pmatrix},$$

$$\mathbf{p}_n^+ = \begin{pmatrix} p_{n0}^{(10)} & p_{n0}^{(01)} & 0 & \ddots & 0 & 0 \\ 0 & p_{n-11}^{(10)} & p_{n-11}^{(01)} & 0 & \ddots & 0 \\ \ddots & 0 & p_{n-22}^{(10)} & p_{n-22}^{(01)} & \ddots & \ddots \\ 0 & \ddots & \ddots & \ddots & \ddots & 0 \\ 0 & 0 & \ddots & 0 & p_{0n}^{(10)} & p_{0n}^{(01)} \end{pmatrix},$$

$$\mathbf{p}_n^- = \begin{pmatrix} p_{n0}^{(-10)} & 0 & \ddots & 0 \\ p_{n-11}^{(0-1)} & p_{n-11}^{(-10)} & \ddots & \ddots \\ 0 & p_{n-22}^{(0-1)} & \ddots & 0 \\ \ddots & \ddots & \ddots & p_{1n-1}^{(-10)} \\ 0 & \ddots & 0 & p_{0n-1}^{(0-1)} \end{pmatrix}.$$

Finally, we introduce the column vector

$$\mathbf{C}_n(t) = \begin{pmatrix} \mathbf{c}_{2n-1}(t) \\ \mathbf{c}_{2n}(t) \end{pmatrix},$$

so that Eq. (39) may be written

$$\tau_D \dot{\mathbf{C}}_n(t) = \mathbf{Q}_n^- \mathbf{C}_{n-1}(t) + \mathbf{Q}_n \mathbf{C}_n(t) + \mathbf{Q}_n^+ \mathbf{C}_{n+1}(t) + \xi_\ell(t) \left[ \mathbf{P}_n^- \mathbf{C}_{n-1}(t) + \mathbf{P}_n \mathbf{C}_n(t) + \mathbf{P}_n^+ \mathbf{C}_{n+1}(t) \right], \quad (41)$$

where

$$\mathbf{Q}_n^\pm = \begin{pmatrix} \mathbf{q}_{2n-1}^{\pm\pm} & \mathbf{0} \\ \mathbf{0} & \mathbf{q}_{2n}^{\pm\pm} \end{pmatrix}, \quad \mathbf{Q}_n = \begin{pmatrix} \mathbf{q}_{2n-1} & \mathbf{0} \\ \mathbf{0} & \mathbf{q}_{2n} \end{pmatrix}, \quad (42)$$

$$\mathbf{P}_n^- = \begin{pmatrix} \mathbf{0} & \mathbf{p}_{2n-1}^- \\ \mathbf{0} & \mathbf{0} \end{pmatrix}, \quad \mathbf{P}_n = \begin{pmatrix} \mathbf{0} & \mathbf{p}_{2n-1}^+ \\ \mathbf{p}_{2n}^- & \mathbf{0} \end{pmatrix}, \quad \mathbf{P}_n^+ = \begin{pmatrix} \mathbf{0} & \mathbf{0} \\ \mathbf{p}_{2n}^+ & \mathbf{0} \end{pmatrix}. \quad (43)$$

The set of Eqs. (41) lends itself easily to perturbation theory, and are furthermore tridiagonal, so that the resulting equations can be solved via the matrix continued fraction method [40]. Thus, we proceed by setting

$$b_{n,m}(t) = b_{n,m}^{(0)} + b_{n,m}^{(1)}(t), \tag{44}$$

where the superscript $^{(i)}$ indicates the order of the field strength. It naturally follows that

$$\mathbf{C}_n(t) = \mathbf{C}_n^{(0)} + \mathbf{C}_n^{(1)}(t), \tag{45}$$

so that Eq. (41) is immediately transformed into the vector perturbation equations

$$\mathbf{Q}_n^- \mathbf{C}_{n-1}^{(0)} + \mathbf{Q}_n \mathbf{C}_n^{(0)} + \mathbf{Q}_n^+ \mathbf{C}_{n+1}^{(0)} = \mathbf{0} \tag{46}$$

and

$$\tau_D \dot{\mathbf{C}}_n^{(1)}(t) = \mathbf{Q}_n^- \mathbf{C}_{n-1}^{(1)}(t) + \mathbf{Q}_n \mathbf{C}_n^{(1)}(t) + \mathbf{Q}_n^+ \mathbf{C}_{n+1}^{(1)}(t) + \xi_\ell(t) \mathbf{K}_n^{(0)}, \tag{47}$$

where

$$\mathbf{K}_n^{(0)} = \mathbf{P}_n^- \mathbf{C}_{n-1}^{(0)}(t) + \mathbf{P}_n \mathbf{C}_n^{(0)} + \mathbf{P}_n^+ \mathbf{C}_{n+1}^{(0)}$$

Since we are interested to the linear response to a sinusoidal AC field, we can set

$$\xi_\ell(t) = \tilde{\xi}_\ell(\omega) e^{i\omega t}, \quad \mathbf{C}_n^{(1)}(t) = \tilde{\xi}_\ell(\omega) \tilde{\mathbf{C}}_n^{(1)}(\omega) e^{i\omega t} \tag{48}$$

in Eq. (47), so that this equation finally becomes

$$i\omega \tau_D \tilde{\mathbf{C}}_n^{(1)}(\omega) = \mathbf{Q}_n^- \tilde{\mathbf{C}}_{n-1}^{(1)}(\omega) + \mathbf{Q}_n \tilde{\mathbf{C}}_n^{(1)}(\omega) + \mathbf{Q}_n^+ \tilde{\mathbf{C}}_{n+1}^{(1)}(\omega) + \mathbf{K}_n^{(0)}. \tag{49}$$

Now, Eqs. (46) and (49) are algebraic vector equations that may be solved. By introducing the matrix continued fraction

$$\boldsymbol{\Delta}_n(s) = \left[ s\mathbf{I} - \mathbf{Q}_n - \mathbf{Q}_n^+ \boldsymbol{\Delta}_{n+1}(s)\mathbf{Q}_{n+1}^- \right]^{-1} \tag{50}$$

and the notation

$$\mathbf{S}_n(s) = \boldsymbol{\Delta}_n(s)\mathbf{Q}_n^-, \tag{51}$$

however, the *exact* solution of Eqs. (46) and (49) can be written down in terms of $\mathbf{S}_n(s)$, $\boldsymbol{\Delta}_n(s)$ and $\mathbf{K}_n^{(0)}$ only [19,40]. In particular, we have

$$\mathbf{C}_n^{(0)} = \mathbf{S}_n(0)\mathbf{S}_{n-1}(0)\ldots\mathbf{S}_1(0), \ n > 1, \tag{52}$$

$$\mathbf{C}_1^{(0)} = \mathbf{S}_1(0), \tag{53}$$

$$\tilde{\mathbf{C}}_n^{(1)}(\omega) = \boldsymbol{\Delta}_n(i\omega\tau_D)\left\{\mathbf{Q}_n^-\tilde{\mathbf{C}}_{n-1}^{(1)}(\omega) + \mathbf{K}_n^{(0)} + \sum_{p=1}^{\infty}\prod_{k=1}^{p}\mathbf{Q}_{n+k-1}^+\boldsymbol{\Delta}_{n+k}(i\omega\tau_D)\mathbf{K}_{n+p}^{(0)}\right\}, n > 1 \tag{54}$$

$$\tilde{\mathbf{C}}_1^{(1)}(\omega) = \boldsymbol{\Delta}_1(i\omega\tau_D)\left\{\mathbf{K}_1^{(0)} + \sum_{p=1}^{\infty}\prod_{k=1}^{p}\mathbf{Q}_k^+\boldsymbol{\Delta}_{k+1}(i\omega\tau_D)\mathbf{K}_{p+1}^{(0)}\right\} \tag{55}$$

Thus, Eqs. (52)-(55) lead to the exact steady-state solution of Eq. (36) in terms of matrix continued fractions, which in turn will yield the solution of Eq. (2) in the linear regime, demonstrated in the next section.

## V. DYNAMICAL EQUATION OF STATE FOR INTERACTING DIPOLES

Here we first develop an exact solution of Eq. (2) from which we will deduce an exact equation for the complex linear polarization response obtained *from molecular considerations*. We start from Eqs. (2) and (30), and expand $W(\mathbf{u},t)$ in series of Legendre polynomials, viz.

$$W(\mathbf{u},t) = \sum_{n=0}^{\infty}\left(n + \frac{1}{2}\right)a_n(t)P_n(\cos\vartheta)$$

so that accounting for Eq. (37) we have

$$\frac{2\tau_D}{n(n+1)}\dot{a}_n(t)+a_n(t)=\frac{\xi_\ell(t)}{(2n+1)}\big(a_{n-1}(t)-a_{n+1}(t)\big)\mp\frac{\lambda}{(2n+1)}\big(b_{n+1,1}(t)-b_{n-1,1}(t)\big). \qquad (56)$$

Now Eq. (1) can be written

$$P(t)=\rho_0\mu a_1(t), \qquad (57)$$

Thus we are strictly interested in the steady-state solution of Eq. (56) for $n=1$. This equation is

$$\tau_D\dot{a}_1(t)+a_1(t)=\frac{\tilde{\xi}_\ell(\omega)}{3}e^{i\omega t}\big(1-a_2(t)\big)\mp\frac{\lambda}{3}\big(b_{2,1}(t)-b_{0,1}(t)\big) \qquad (58)$$

We may solve Eq. (58) by perturbation theory just like in the preceding section. We set

$$a_n(t)=a_n^{(0)}+\tilde{\xi}_\ell(\omega)A_n^{(1)}(\omega)e^{i\omega t},\ b_{n,m}(t)=b_{n,m}^{(0)}+\tilde{\xi}_\ell(\omega)B_{n,m}^{(1)}(\omega)e^{i\omega t}, \qquad (59)$$

so that

$$a_1^{(0)}=\mp\frac{\lambda}{3}\big(b_{2,1}^{(0)}-b_{0,1}^{(0)}\big)=0, \qquad (60)$$

as it must, since this coefficient determines the static polarization in the absence of external field, and

$$a_2^{(0)}=\mp\frac{\lambda}{5}\big(b_{3,1}^{(0)}-b_{1,1}^{(0)}\big)\neq 0. \qquad (61)$$

Because of Eq. (57), the steady state linear polarization is

$$P(t)=\rho_0\mu\tilde{\xi}_\ell(\omega)A_1^{(1)}(\omega)e^{i\omega t} \qquad (62)$$

where Eq. (59) has been used. Using Eqs. (58), (59) and (61), we have

$$A_1^{(1)}(\omega)=\frac{1}{3(1+i\omega\tau_D)}\left\{1\pm\frac{\lambda}{5}\big(b_{3,1}^{(0)}-b_{1,1}^{(0)}\big)\mp\lambda\big(B_{2,1}^{(1)}(\omega)-B_{0,1}^{(1)}(\omega)\big)\right\} \qquad (63)$$

This result is central to our work. For purely polar molecules, $\tilde{\xi}_\ell(\omega)$ is given by [4]

$$\tilde{\xi}_\ell(\omega) = \frac{3\varepsilon(\omega)\xi_M}{2\varepsilon(\omega)+1}, \tag{64}$$

where $\xi_M = \beta\mu E$ and $E$ is the amplitude of the Maxwell field. By using the macroscopic definition of the polarization in terms of that field, viz.

$$4\pi P(t) = (\varepsilon(\omega)-1)Ee^{i\omega t}, \tag{65}$$

we have, equating the right hand sides of Eqs. (62) and (64) the dynamical equation of state for purely polar dielectrics, viz.

$$\frac{(\varepsilon(\omega)-1)(2\varepsilon(\omega)+1)}{3\varepsilon(\omega)} = 3\lambda A_1^{(1)}(\omega) = \frac{4\pi\alpha(\omega)}{\Upsilon}, \tag{66}$$

where the linear dynamic polarizability $\alpha(\omega)$ is given by

$$\alpha(\omega) = \beta N \mu^2 A_1^{(1)}(\omega) \tag{67}$$

and where $N$ is the number of molecular dipoles contained in the specimen of volume $\Upsilon$ under consideration.

It is also possible to obtain an analytical representation for the complex polarizability using the formula described in Ref.19, Chapter 2 for the low frequency linear response of systems governed by Fokker-Planck equations. This formula is

$$\frac{B_{n,1}^{(1)}(\omega)}{B_{n,1}^{(1)}(0)} = \frac{\Delta_{n,1}}{1+i\omega/\Lambda_1} + \frac{1-\Delta_{n,1}}{1+i\omega\tau_{n,1}^W} \tag{68}$$

where $\Lambda_1$ is the smallest non-vanishing eigenvalue of the Fokker-Planck equation (36), and [19, 39]

$$\Delta_{n,1} = \frac{\tau_{n,1}/\tau_{n,1}^{eff} - 1}{\Lambda_1 \tau_{n,1} - 2 + \left(\Lambda_1 \tau_{n,1}^{eff}\right)^{-1}}, \tag{69}$$

$$\tau_{n,1}^{W} = \frac{\Lambda_1 \tau_{n,1} - 1}{\Lambda_1 \tau_{n,1}^{eff} - 1} \tau_{n,1}^{eff} \tag{70}$$

$$\tau_{n,1} = \lim_{\omega \to 0} \frac{B_{n,1}^{(1)}(\omega) - B_{n,1}^{(1)}(0)}{i\omega B_{n,1}^{(1)}(0)} \tag{71}$$

$$\tau_{n,1}^{eff} = \lim_{\omega \to \infty} \frac{B_{n,1}^{(1)}(0)}{i\omega B_{n,1}^{(1)}(\omega)} \tag{72}$$

Equation (68) shows that the response of a pair of dipoles consists of two relaxation processes, namely one Kramers-like thermally activated inter-well process onto which is superposed an (overdamped) intra-well relaxation process. By substituting Eq. (68) into Eq. (63), one may see that the relaxation dynamics of the polarization is governed by not less than *four* relaxation time scales in the idealized picture of a real system that we have chosen here. These time scales are $\tau_D$, the free rotational diffusion time, $\Lambda_1^{-1}$, the time scale for thermally activated reversal of the dipoles (inter-well process) and $\tau_{0,1}^{W}$ and $\tau_{2,1}^{W}$. The two latter time scales are related to the same phenomenon, i.e. relaxation of the dipolar motion *inside* the two potential wells (intra-well process) [19].

## VI. RESULTS AND DISCUSSION

The analytical formula that we have obtained for the dynamic permittivity through Eqs. (63) and (66) is a simple one in terms of physical interpretation : the polarizability of the sample Eq. (63) is made of a superposition of an ideal gas part and a part due to intermolecular interactions. Then, the second degree Eq. (66) is solved in such a way that the retained root that is positive at zero frequency. The results of the computations of Eq. (63) are given on Figures 1 ( $g_K > 1$ ) and 2 ( $g_K < 1$ ).

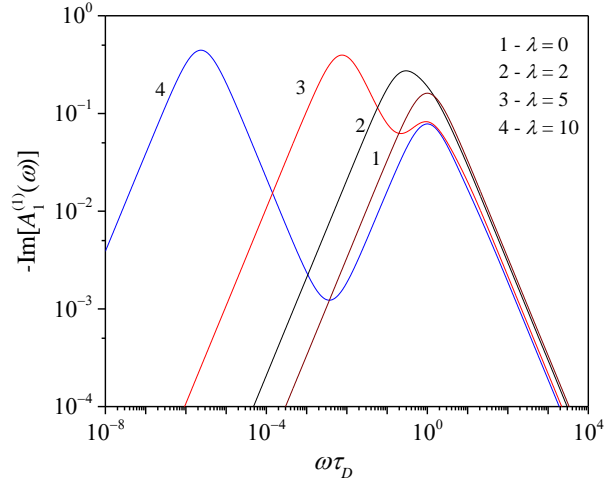

Figure 1 (color online) : the imaginary part of Eq. (63) as a function of frequency for various values of $\lambda$ for $g_K > 1$.

For $g_K > 1$, when $\lambda$ is increased, the dielectric spectra evolve from a single Lorentzian having its maximum at $\omega = 1/\tau_D$ to two separate Lorentzians, one of them peaking at $\omega = \Lambda_1$ and the other having a maximum at $\omega = 1/\tau_D$. Hence, at large $\lambda$ values, the amplitudes of the time scales $\tau_{n,1}^W$ are masked by the contribution of the Debye peak, because $\Delta_{n,1} \approx 1$. Therefore, at large values of the interaction parameter $\lambda$, the peak of lowest frequency may be associated with a collective mode (synchronous reversal of a pair of strongly coupled dipoles), while the peak of frequency $\omega = 1/\tau_D$ may, to some extent (because the amplitude of this peak depends on $\lambda$), be associated with single molecular motion since $\tau_D$ is the relaxation time scale of a molecule in the absence of long range interactions.

On the contrary, for $g_K < 1$, there is effectively only one peak. This peak first shifts to *higher* frequencies and its amplitude flattens as the parameter $\lambda$ increases, as illustrated in Figure 2. Then,

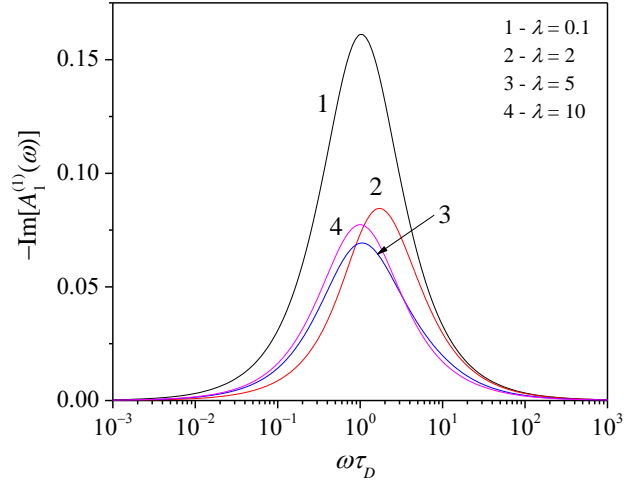

Figure 2 (color online) : the imaginary part of Eq. (63) as a function of frequency for various values of $\lambda$ for $g_K < 1$.

the peak shifts back to the value $\omega = 1/\tau_D$, and its amplitude increases again. Thus, the dynamics reflects that of an effective single molecule, where quasi-static interactions have no effect on the relaxation characteristic time scales.

Next, it is possible to compute $\Lambda_1$ with the help of the matrix continued fraction method. More precisely, this quantity can be computed as the smallest non-vanishing root of the equation [40]

$$\det\left[\Lambda \mathbf{I} + \mathbf{Q}_1 + \mathbf{Q}_1^+ \mathbf{\Delta}_2(-\Lambda)\mathbf{Q}_2^-\right] = 0. \qquad (73)$$

In practice, Eq. (73) is solved using the Newton-Raphson method. Not more than 5 iterations of the Newton-Raphson algorithm are necessary to obtain numerical convergence in the majority of the cases.

Furthermore, $\Lambda_1$ has Arrhenius-Kramers behavior. This is shown in Figure 3, where $\Lambda_1$ is plotted as a function of $\lambda$. By using the Kramers-Langer [17, 41] theory, the asymptotic behavior of $\Lambda_1$ is given by

$$\Lambda_1 \tau_D \approx 64 \left( \frac{\lambda}{\pi} \right)^{\frac{5}{2}} e^{-2\lambda}, \quad \lambda \gg 1 \tag{74}$$

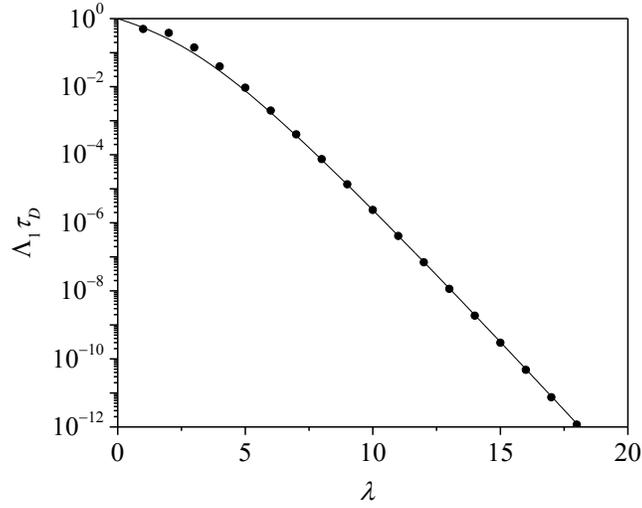

Figure 3 : The smallest non-vanishing eigenvalue of the Fokker-Planck equation (36). Solid line : Numerical calculation from Eq.(73). Dots : Eq.(74).

Clearly, Eq. (74) yields an excellent representation of $\Lambda_1$ for large $\lambda$. For small $\lambda$, we have instead

$$\Lambda_1 \tau_D \approx 1 \mp \frac{\lambda}{3}, \quad \lambda \ll 1, \tag{75}$$

the two signs corresponding to $g_K > 1$ and $g_K < 1$ respectively. The preceding analysis clearly shows the absence of thermally activated behavior for $g_K < 1$ because the two equilibrium states are those of anti-parallel alignment $(0, \pi)$ and $(\pi, 0)$, which cannot be distinguished since the system we consider here is made of identical molecules, i.e. for $g_K < 1$, we have $\Delta_{n,1} \approx 0$. This is so in spite of the fact that Eq. (73) yields the same smallest non-vanishing eigenvalue for both signs in Eq. (30) at large $\lambda$. This is in contrast with the situation for $g_K > 1$ where a pair of dipoles have two distinct equilibrium orientations given by $(0,0)$ and $(\pi,\pi)$, and where here indeed, a thermally activated process takes place. Thus, we may write

$$\frac{B_{n,1}^{(1)}(\omega)}{B_{n,1}^{(1)}(0)} \approx \frac{1}{1+i\omega/\Lambda_1} \tag{76}$$

for $g_K > 1$ and

$$\frac{B_{n,1}^{(1)}(\omega)}{B_{n,1}^{(1)}(0)} \approx \frac{1}{1+i\omega\tau_{n,1}^W} \tag{77}$$

for $g_K < 1$. In the latter situation, one may demonstrate using the effective eigenvalue method [19] that we have

$$\tau_{n,1}^W \approx \tau_D \left(1+\frac{\lambda}{3}\right)^{-1} \tag{78}$$

and from our numerical results this formula is largely sufficient for all values of $\lambda$. Equations (63), (74)-(78) allow us to obtain two analytical formulas for $A_1^{(1)}(\omega)$, according to the value of $g_K$. These are

$$A_1^{(1)}(\omega) \approx \frac{1+\lambda\left(b_{3,1}^{(0)}-b_{1,1}^{(0)}\right)/5}{3(1+i\omega\tau_D)} + \frac{\lambda}{3}\frac{B_{0,1}^{(1)}(0)-B_{2,1}^{(1)}(0)}{(1+i\omega\tau_D)(1+i\omega/\Lambda_1)} \tag{79}$$

for $g_K > 1$, and

$$A_1^{(1)}(\omega) \approx \frac{1-\lambda\left(b_{3,1}^{(0)}-b_{1,1}^{(0)}\right)/5}{3(1+i\omega\tau_D)} - \frac{\lambda}{3}\frac{B_{0,1}^{(1)}(0)-B_{2,1}^{(1)}(0)}{(1+i\omega\tau_D)\left(1+\dfrac{i\omega\tau_D}{1+\lambda/3}\right)} \tag{80}$$

for $g_K < 1$. In Eq. (79), $\Lambda_1$ may safely be replaced by Eq. (74) for $\lambda > 10$. Comparison of Eqs. (79) and (80) with the exact solution computed with the help of matrix continued fractions are shown on Figures 4 and 5. Clearly, Eqs. (79) and (80) represent $A_1^{(1)}(\omega)$ fairly well in the whole frequency range for all values of $\lambda$, demonstrating the correctness of our analysis in terms of significant eigenmodes of the Fokker-Planck equation (36).

The complex susceptibility can also be calculated analytically from Eq. (66). It is given by

$$4\pi\chi(\omega) = \frac{1}{4}\left[9\lambda A_1^{(1)}(\omega) - 3 + \sqrt{8 + \left(1 + 9\lambda A_1^{(1)}(\omega)\right)^2}\right], \tag{81}$$

resulting in two expressions for this quantity, one for $g_K > 1$ and one for $g_K < 1$. Furthermore, from Eq. (66), we have

$$\frac{\chi(\omega)}{\chi'(0)}G(\omega) = \frac{A_1^{(1)}(\omega)}{A_1^{(1)}(0)} = \frac{\alpha(\omega)}{\alpha'(0)}, \tag{82}$$

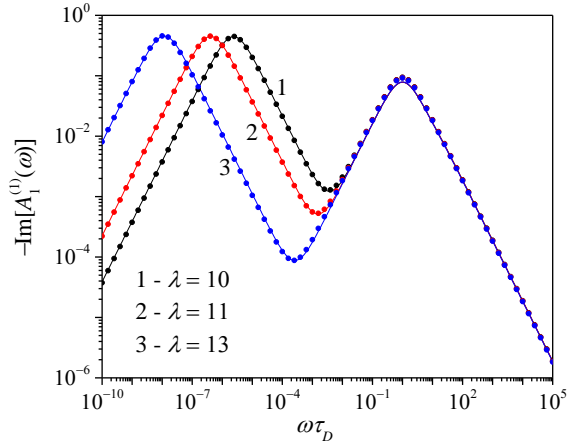

Figure 4 (color online) : Imaginary part of $A_1^{(1)}(\omega)$ vs $\omega\tau_D$ for several values of $\lambda$ that are relevant to polar fluids. Solid line: Eq.(63) for $g_K > 1$ (upper sign). Dots : Eq.(79).

where we have introduced a reduced dynamical internal field factor $G(\omega)$ defined by

$$G(\omega) = \frac{1+(2\varepsilon(\omega))^{-1}}{1+(2\varepsilon'(0))^{-1}} = G'(\omega) + iG''(\omega). \tag{83}$$

As we have seen before, the thermally activated process occurs for $g_K > 1$ only. By plotting $G'(\omega)$ and $G''(\omega)$ at $\lambda = 10$ using Eq. (81) (i.e. a value at which the thermally activated process is well resolved in frequency), so that in the frequency region of the thermally activated process $G'(\omega)$ does not seriously depart from unity while $G''(\omega)$ is vanishingly small. This is illustrated in Figure 6.

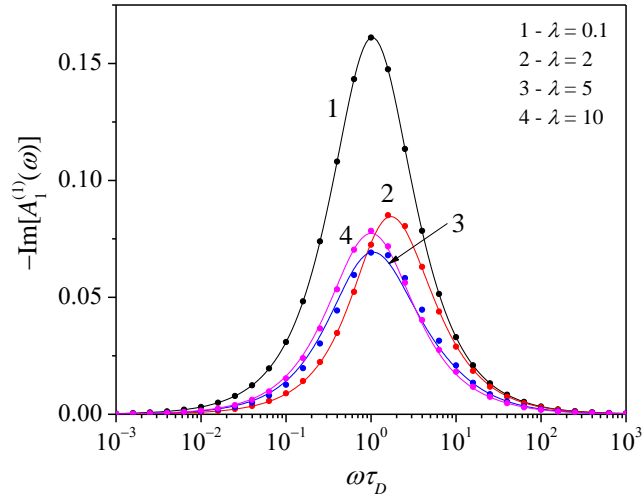

Figure 5 (color online) : $\mathrm{Im}\left[A_1^{(1)}(\omega)\right]$ vs $\omega\tau_D$ for various values of $\lambda$. Solid line : Eq. (63) for $g_K < 1$. Dots : Eq.(80).

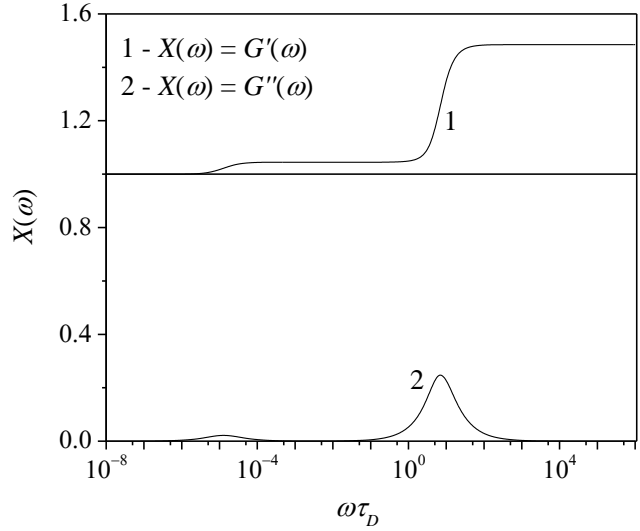

Figure 6 : The real and imaginary parts of $G(\omega)$ vs $\omega\tau_D$ for $\lambda = 10$.

Therefore, in the frequency range of the thermally activated process, to a very good approximation

$$\frac{\chi(\omega)}{\chi'(0)} \approx \frac{\alpha(\omega)}{\alpha'(0)}. \tag{84}$$

where we have introduced the dielectric complex susceptibility via $\chi(\omega) = (\varepsilon(\omega)-1)/(4\pi)$. This implies that to analyze the thermally activated process, *there is no need to account for dynamical internal field effects.* At higher frequencies, this analysis is more difficult to perform analytically. This is why we construct a Cole-Cole plot for $\chi(\omega)/\chi'(0)$ and superimpose that of $\alpha(\omega)/\alpha'(0)$ for $g_K > 1$ for $\lambda = 10$ on the same Figure. The result of this superposition is shown on Figure 7.

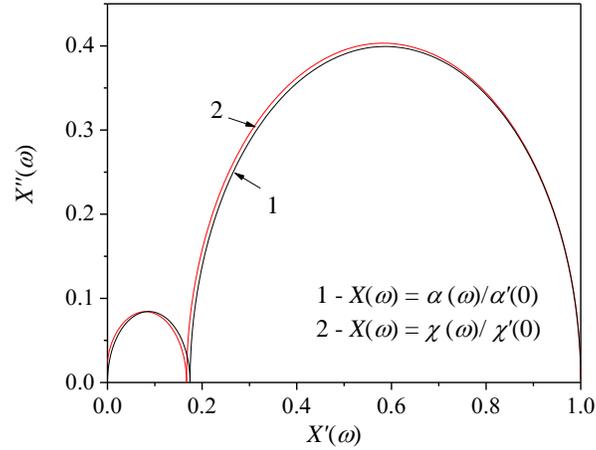

Figure 7 (color online) : Cole-Cole plots of the susceptibility and the polarizability of polar fluids for $\lambda = 10$ and $g_K > 1$.

As frequency increases, slight deviations occur between $\dfrac{\chi(\omega)}{\chi'(0)}$ and $\dfrac{\alpha(\omega)}{\alpha'(0)}$, however, these are less than 5% in relative error, and therefore, we may state that *up to that relative theoretical error*, Eq. (84) holds at all frequencies beyond the resonant ones, so that the *dynamical effect* of the internal field can be neglected at the lowest approximation order. This is indeed not so at the static level.

## VII.    COMPARISON WITH PREVIOUS THEORETICAL RESULTS

Now, we compare our results with those obtained in Refs. [28, 29], which are relevant to single-domain ferromagnetic particles, but which we believe to be relevant for long range interacting dipole systems in general, and therefore, to polar fluids also. In Refs. [28, 29], the preferred parallel alignment was handled, however in their approach $g_K \approx 1$ since this is a modified mean field approach. There, the static susceptibility (in our notation) was found to be

$$4\pi\chi \approx \lambda\left(1+\frac{\lambda}{3}+\frac{\lambda^2}{144}+...\right) \tag{85}$$

and in this approach, clearly the cubic term is negligible since $\lambda << 1$, so that it contributes nothing. In our approach where correlations are accounted for, we find in the same context before internal field corrections the formula for the complex polarizability

$$4\pi\alpha \approx \lambda\left(1+\frac{\lambda}{3}+\frac{\lambda^2}{15}+...\right), \quad g_K > 1, \lambda << 1 \tag{86}$$

and after internal field corrections, we find

$$4\pi\chi \approx \lambda\left(1+\frac{2\lambda}{3}+\frac{8\lambda^2}{45}+...\right). \tag{87}$$

This means that even at very low concentrations, the *static* effects due to the internal field on the static susceptibility cannot be ignored. Including internal field effects, the susceptibility derived in Ref. 28 becomes

$$4\pi\chi \approx \lambda\left(1+\frac{2\lambda}{3}+\frac{17\lambda^2}{144}+...\right) \tag{88}$$

so that the formalism derived in Ref. 28 agrees in all respects with the present one provided $\lambda << 1$, since the cubic terms of Eqs. (87) and (88) are nearly the same numerically. Comparison for higher values of $\lambda$ cannot be carried out since the results presented in Ref. 28 are valid for low $\lambda$ only. These formulas are however to be compared with Onsager's formula for which correlations are totally neglected, viz.

$$4\pi\chi_O \approx \lambda\left(1+\frac{\lambda}{3}-\frac{\lambda^2}{9}+...\right), \quad g_K = 1, \lambda << 1, \tag{89}$$

emphasizing that $g_K = 1$ (therefore a temperature-independent Kirkwood correlation factor) is a special case that still has to be worked out [34]. It has to be remarked that Eq. (87) is also in agreement with the approach developed in Ref. 31, where the standard mean field approach was developed and where the applicability to real systems was not possible because of the existence of a Curie point. As expected, the formalism presented here removes this point due to inclusion of orientational correlations. To conclude with static internal field effects, we have derived large $\lambda$ approximations for the static susceptibility. These are

$$4\pi\chi(\lambda) \approx \frac{21}{4}\lambda - \frac{133}{32}, \quad \lambda \gg 1 \tag{90}$$

for $g_K > 1$ and

$$4\pi\chi(\lambda) \approx \frac{3}{4}\lambda - \frac{11}{16}, \quad \lambda \gg 1 \tag{91}$$

for $g_K < 1$. These equations can be applied for $\lambda > 4$. They show the static effect of the internal field in that the susceptibility is grossly five times that of the ideal gas of dipoles for $g_K > 1$ and 75% of the susceptibility of the ideal gas for $g_K < 1$.

The dynamical approach to the magnetic susceptibility developed in Ref. 29 leads to the analytical expressions for the real and imaginary parts

$$4\pi\chi'(\omega) = \frac{\lambda}{1+\omega^2\tau_D^2}\left(1 + \frac{\lambda(1-\omega^2\tau_D^2)}{3(1+\omega^2\tau_D^2)}\right), \tag{92}$$

$$4\pi\chi''(\omega) = \frac{\lambda\omega\tau_D}{1+\omega^2\tau_D^2}\left(1 + \frac{2\lambda}{3(1+\omega^2\tau_D^2)}\right). \tag{93}$$

Our results agree with Eqs. (92) and (93) for $\lambda \ll 1$ since the second term in between braces in the right hand side of these equations is small. Furthermore, the leading term in these equations is exactly the mean field one [42]. The second term in the right hand side of Eqs. (92) and (93) has similar behavior with the one found in Eq. (63) at low $\lambda$ values therefore, both results are in qualitative agreement again for $\lambda \ll 1$.

At large $\lambda$ and for $g_K > 1$, a thermally activated process sets in. A scaling law for the barrier occurring in the expression for the relaxation time of glass forming liquids was suggested by Tarjus et al. [43] in the form

$$\ln \tau = \ln \tau_0 + \beta E_\infty(\rho_0) \Phi\left(\left(\beta E_\infty(\rho_0)\right)^{-1}\right) \tag{94}$$

where $\tau_0$ is a pre-exponential factor, $E_\infty(\rho_0)$ is the energy barrier away from the glass transition temperature $T_g$ and $\Phi(z)$ a monotonically decreasing function of $z$ that scales the effective activation energy related to the $\alpha$ process and which decreases to unity at large $z$. This scaling law is valid for isochoric situations, and was found in agreement with numerical simulations. Of course, our present calculations are unable to reproduce Eq. (94), mainly because our premature "truncation" of the BBGKY process generated by the Dean-Kawasaki equation. Yet, from Eq. (74), we suggest that if such a scaling law could be used for dielectric relaxation, then we may write $E_\infty(\rho_0) = C\lambda$, where $C$ is a numerical factor that depends of the kind of interaction considered in the liquid phase, since our calculations are clearly not valid near $T_g$. In the example treated here, $C = 2$ while in the framework of other intermolecular interactions, $C$ might be different [44].

At last, a word concerning the full dipole-dipole interaction is necessary. On taking the inter-particle distance along the Z axis, the long range interaction potential is, in our notation

$$\beta U_m(\mathbf{u},\mathbf{u}') = -2\lambda\cos\vartheta\cos\vartheta' + \lambda\sin\vartheta\sin\vartheta'\cos(\varphi-\varphi')$$

where $\varphi$ and $\varphi'$ are the azimuthal angular coordinates of the dipoles with respective orientations $\mathbf{u}$ and $\mathbf{u}'$. This results in the effective pair interaction potential

$$\beta V_2^{eff}(\mathbf{u},\mathbf{u}') = -2\lambda\cos\vartheta\cos\vartheta' + \lambda\sin\vartheta\sin\vartheta'\cos(\varphi-\varphi') - 3\lambda\cos^2\vartheta - 3\lambda\cos^2\vartheta'. \qquad (95)$$

This effective potential exhibits minima separated by potential barriers. Therefore according to the Kramers theory, a thermally activated process will exist in this case too. Hence, our results will be qualitatively unchanged.

## VIII. CONCLUSION

In this work, we have derived an analytical formula for the dynamic susceptibility of dipolar assemblies beyond the mean field approximation. In order to accomplish this, we have suggested a way of stopping the BBGKY-like process generated by the Dean-Kawasaki formalism, suitably adapted to rotational Brownian motion of long range interacting molecules by Cugliandolo et al. [33]. The main results of our work are that we could derive our formula (63) beyond the mean field approximation which allowed us in turn to derive an analytical formula for the complex susceptibility of interacting dipoles with Eq. (81). There it was shown that if long range intermolecular interactions are accounted for beyond the mean field approximation, then such interactions may induce a thermally activated process in the dynamics provided the Kirkwood correlation factor $g_K$ is larger than unity. In the opposite situation $g_K < 1$, no thermally activated process arises and the position of the dipolar low-frequency absorption peak is practically unaffected. Typical values for $\lambda$ for polar liquids at room temperature are around 10, while for magnetic nanoparticles particles, $\lambda$ is between 5 and 8 (for Co and Fe nanoparticles respectively) having diameter around 20 nm (slightly above the critical size for single-domain behavior), and goes down to values at most 0.5 for particles having 8 nm in diameter (Co and Fe have very high bulk saturation magnetization), so that the various formulas derived in this work may be of some use in such nanoparticle assemblies. The magneto-crystalline anisotropy of the particles together with local demagnetizing effects has been ignored here. Yet, we believe that the relevant relaxation time would be of the form $\tau \approx \tau_0 \exp(\sigma + C\lambda)$ for $g_K > 1$ and $\tau \approx \tau_0 \exp(\sigma)$ for $g_K < 1$, where $\tau_0$ is a pre-exponential factor, $C$ a numerical constant and $\sigma$ is the anisotropy to thermal energy ratio. In particular, the above result for $g_K > 1$ validates the empirical procedure used in Refs. 20 and 24 for the estimation of the relaxation time of magnetic nanoparticle assemblies.

It may further be possible that in such assemblies, $\lambda$ as defined in this work would render values that are too small to allow consistent contribution of dipole-dipole interactions. Thus, if inter-particle interactions are to play a role on the dynamic susceptibility beyond these sizes, another mechanism than the dipole-dipole interaction has to be found in order to explain experimental data.

## ACKNOWLEDGMENTS

We would like to thank a number of individuals who have helped and encouraged us in the course of this work. Particularly, we would like to warmly thank Profs. F. van Wijland, Yu. L. Raikher, Yu. P. Kalmykov, and Drs. F. Ladieu and C. Alba-Simionesco for helpful discussions and useful comments.

## APPENDIX A : EXPLICIT EXPRESSIONS FOR THE COEFFICIENTS IN THE DIFFERENTIAL RECURRENCE EQS. (38)

Here we give explicit expressions for the coefficients that are involved in Eqs. (38). These are

$$q_{nm}^{(00)} = -\frac{n(n+1)}{2}\left[1 - \frac{\lambda}{(2n-1)(2n+3)}\right] - \frac{m(m+1)}{2}\left[1 - \frac{\lambda}{(2m-1)(2m+3)}\right],$$

$$q_{nm}^{(-20)} = \frac{\lambda(n-1)n(n+1)}{2(2n-1)(2n+1)}, \quad q_{nm}^{(20)} = -\frac{\lambda n(n+1)(n+2)}{2(2n+1)(2n+3)},$$

$$q_{nm}^{(0-2)} = \frac{\lambda(m-1)m(m+1)}{2(2m-1)(2m+1)}, \quad q_{nm}^{(02)} = -\frac{\lambda m(m+1)(m+2)}{2(2m+1)(2m+3)},$$

$$q_{nm}^{(-1-1)} = \pm\frac{\lambda mn(m+n+2)}{2(2n+1)(2m+1)}, \quad q_{nm}^{(11)} = \mp\frac{\lambda(n+1)(m+1)(n+m)}{2(2n+1)(2m+1)},$$

$$q_{nm}^{(1-1)} = \pm\frac{\lambda m(n+1)(m-n+1)}{2(2n+1)(2m+1)}, \quad q_{nm}^{(-11)} = \mp\frac{\lambda n(m+1)(m-n-1)}{2(2n+1)(2m+1)},$$

$$p_{nm}^{(\pm 10)} = \mp\frac{n(n+1)}{2(2n+1)}, \quad p_{nm}^{(0\pm 1)} = \mp\frac{m(m+1)}{2(2m+1)}.$$

**REFERENCES**

[1] P. Debye, *Polar Molecules* (Chem. Catalog. Co., New York, 1929; Reprinted Dover, New York, 1954).

[2] J. G. Kirkwood, J. Chem. Phys. **7**, 911 (1939).

[3] H. Fröhlich, *Theory of Dielectrics*, 2nd Edition (Oxford University Press, 1958).

[4] B.K.P. Scaife, *Principles of Dielectrics*, 2nd Edition (Clarendon, Oxford, 1998).

[5] R. Zwanzig, J. Chem. Phys. **38**, 2766 (1963).

[6] T.W. Nee and R. Zwanzig, J. Chem. Phys. **52**, 6353 (1970).

[7] R. Zwanzig, J. Chem. Phys. **38**, 1605 (1963).

[8] R. H. Cole, J. Chem. Phys. **6**, 385 (1938).

[9] E. Fatuzzo and P. R. Mason, Proc. Phys. Soc. (London) **94**, 729 (1967) ; Proc. Phys. Soc. (London) **94**, 741 (1967)

[10] M. W. Evans, G. J. Evans, W. T. Coffey and P. Grigolini, *Molecular Dynamics and the Theory of Broadband Spectroscopy* (Wiley, New York, 1982).

[11] P. Madden and D. Kivelson, Adv. Chem. Phys. **56**, 467 (1984).

[12] R. Fulton, Mol. Phys. **29**, 405 (1975).

[13] B. Bagchi and A. Chandra, Adv. Chem. Phys. **80**, 1 (1991).

[14] J. P. Hansen and I.R. McDonald, *Theory of Simple Liquids*, 3$^{rd}$ Edition (Academic Press, 2006).

[15] L. Néel, Ann. Geophys. **5**, 99 (1949).

[16] W. F. Brown Jr., Phys. Rev. **130**, 1677 (1963).

[17] H. A. Kramers, Physica **7**, 284 (1940).

[18] R. Courant and D. Hilbert, *Methods of Mathematical Physics* (Wiley, Interscience, New York, 1953).

[19] W. T. Coffey and Yu. P. Kalmykov, *The Langevin Equation*, 4$^{th}$ Edition (World Scientific, Singapore, 2017).

[20] J. L. Dormann, D. Fiorani and E. Tronc, Adv. Chem. Phys. **97**, 283 (1997).

[21] S. Bedanta and W. Kleeman, J. Phys. D : Appl. Phys. **42**, 013001 (2009).

[22] J.I. Gittleman, B. Abeles and S. Bozowski, Phys. Rev. B **9**, 3891 (1974).

[23] S. Shtrikman and E. P. Wohlfarth, Phys. Lett. A **85**, 467 (1981).

[24] J. L. Dormann, L. Bessais and D. Fiorani, J. Phys. C **21**, 2015 (1988).

[25] J.O. Andersson, C. Djurberg, T. Jonsson, P. Svedlindh and P. Nordblad, Phys. Rev. B **56**, 13983 (1997).

[26] D. V. Berkov and N. L. Gorn, J. Phys. : Condens. Matter **13**, 4011 (2001).

[27] A. Yu. Zubarev and L. Yu. Iskakova, Phys. Rev. E **63**, 061507 (2001).

[28] A. O. Ivanov and O. B. Kuznetsova, Phys. Rev. E **64**, 041405 (2001).

[29] A. O. Ivanov, V. S. Zverev, and S. S. Kantorovich, Soft Matter **12**, 3507 (2016) ; A. O. Ivanov, S. S. Kantorovich, V. S. Zverev, E. A. Elfimova, A. V. Lebedev and A. F. Pshenichnikov, Phys. Chem. Chem. Phys. **18**, 18342 (2016) ; A. O. Ivanov and V.S. Zverev, Magnetohydrodynamics **52**, 43 (2016).

[30] B. J. Berne, J. Chem. Phys. **62**, 1154 (1975).

[31] P. M. Déjardin, J. Appl. Phys. **110**, 113921 (2011).

[32] K. Kawasaki, Physica A **208**, 35 (1994); D.S. Dean, J. Phys. A : Math. Gen. **29**, L613 (1996).

[33] L. F. Cugliandolo, P. M. Déjardin, G. S. Lozano and F. van Wijland, Phys. Rev. E **91**, 032139 (2015).

[34] P. M. Déjardin, Y. Cornaton, P. Ghesquière, C. Caliot and R. Brouzet, J. Chem. Phys. **148**, (2018).

[35] W. T. Coffey, Adv. Chem. Phys. **63**, 69 (1985).

[36] J. G. Kirkwood, J. Chem. Phys. **3**, 300 (1935).

[37] S. A. Rice and J. Lekner, J. Chem. Phys. **42**, 3559 (1965).

[38] A. Singer, J. Chem. Phys. **121**, 3657 (2004).

[39] N. Wei, P. M. Déjardin, Yu. P. Kalmykov and W. T. Coffey, Phys. Rev. E. **93**, 042208 (2016).

[40] H. Risken, *The Fokker-Planck Equation*, 2$^{nd}$ Edition (Springer Berlin, 1989).

[41] J. S. Langer, Ann. Phys. (New York) **54**, 258 (1969) ; P. Hänggi, P. Talkner and M. Borkovec, Rev. Mod. Phys. **62**, 251 (1990) ; W. T. Coffey, D. A. Garanin and D. J. McCarthy, Adv. Chem. Phys. **117**, 483 (2001).

[42] P. M. Déjardin and F. Ladieu, J. Chem. Phys. **140**, 034506 (2014).

[43] G. Tarjus, D. Kivelson, S. Mossa and C. Alba-Simionesco, J. Chem. Phys. **120**, 6135 (2004).

[44] P.M. Déjardin, unpublished work (2018).